\documentclass[aip,apl,reprint,superscriptaddress
]{revtex4-1}
\usepackage{graphicx}
\usepackage{amsmath,amsfonts,amssymb}
\usepackage{xcolor}
\usepackage{epstopdf}
\usepackage[retainplus]{siunitx}
\usepackage[version=4]{mhchem}
\usepackage{graphicx}
\usepackage{subfigure}
\usepackage{braket}
\usepackage{booktabs}

\newcommand{\red}[1]{\textcolor{black}{#1}} 

\draft 

\begin{document}


\title
  {Experimental and first-principles spectroscopy of Cu$_2$SrSnS$_4$ and Cu$_2$BaSnS$_4$ photoabsorbers}



\author{Andrea Crovetto}
\email[]{Electronic mail: andrea.crovetto@helmholtz-berlin.de}
\affiliation{SurfCat, DTU Physics, Technical University of Denmark, DK-2800 Kgs. Lyngby, Denmark}
\affiliation{Department of Structure and Dynamics of Energy Materials, Helmholtz-Zentrum Berlin f\"ur Materialien und Energie GmbH, Berlin, Germany}

\author{Zongda Xing}
\affiliation{Department of Chemistry, University College London, 20 Gordon Street, London WC1H 0AJ, United Kingdom}
\affiliation{Thomas Young Centre, University College London, Gower Street, London WC1E 6BT, United Kingdom}

\author{Moritz Fischer}
\affiliation{DTU Fotonik, Technical University of Denmark, DK-2800 Kgs. Lyngby, Denmark}
\affiliation{Center for Nanostructured Graphene (CNG), Technical University of Denmark, DK-2800 Kongens~Lyngby, Denmark}

\author{Rasmus Nielsen}
\affiliation{SurfCat, DTU Physics, Technical University of Denmark, DK-2800 Kgs. Lyngby, Denmark}

\author{Christopher N. Savory}
\affiliation{Department of Chemistry, University College London, 20 Gordon Street, London WC1H 0AJ, United Kingdom}
\affiliation{Thomas Young Centre, University College London, Gower Street, London WC1E 6BT, United Kingdom}

\author{Tomas Rindzevicius}
\affiliation{DTU Health Technology, Technical University of Denmark, DK-2800 Kgs. Lyngby, Denmark}

\author{Nicolas Stenger}
\affiliation{DTU Fotonik, Technical University of Denmark, DK-2800 Kgs. Lyngby, Denmark}
\affiliation{Center for Nanostructured Graphene (CNG), Technical University of Denmark, DK-2800 Kongens~Lyngby, Denmark}

\author{David O. Scanlon}
\affiliation{Department of Chemistry, University College London, 20 Gordon Street, London WC1H 0AJ, United Kingdom}
\affiliation{Thomas Young Centre, University College London, Gower Street, London WC1E 6BT, United Kingdom}
\affiliation{Diamond Light Source Ltd., Diamond House, Harwell Science and Innovation Campus, Didcot, Oxfordshire OX11 0DE, United Kingdom}

\author{Ib Chorkendorff}
\affiliation{SurfCat, DTU Physics, Technical University of Denmark, DK-2800 Kgs. Lyngby, Denmark}

\author{Peter C. K. Vesborg}
\affiliation{SurfCat, DTU Physics, Technical University of Denmark, DK-2800 Kgs. Lyngby, Denmark}

\begin{abstract}
The Cu$_2$BaSnS$_4$ (CBTS) and Cu$_2$SrSnS$_4$ (CSTS) semiconductors have been recently proposed as potential wide band gap photovoltaic absorbers. Although several measurements indicate that they are less affected by band tailing than their parent compound Cu$_2$ZnSnS$_4$, their photovoltaic efficiencies are still low.
To identify possible issues, we characterize CBTS and CSTS in parallel by a variety of spectroscopic methods complemented by first-principles calculations.
Two main problems are identified in both materials. The first is the existence of deep defect transitions in low-temperature photoluminescence, pointing to a high density of bulk recombination centers. The second is a low electron affinity, which emphasizes the need for an alternative heterojunction partner and electron contact. We also find a tendency for downward band bending at the surface of both materials. In CBTS, this effect is sufficiently large to cause carrier type inversion, which may enhance carrier separation and mitigate interface recombination. Optical absorption at room temperature is exciton-enhanced in both CBTS and CSTS. Deconvolution of excitonic effects yields band gaps that are about 100~meV higher than previous estimates based on Tauc plots.  Although the two investigated materials are remarkably similar in an idealized, defect-free picture, the present work points to CBTS as a more promising absorber than CSTS for tandem photovoltaics.
\end{abstract}


\maketitle 

\section{Introduction}
Performing charge-neutral, multi-element substitutions in II-VI semiconductors~\cite{Pamplin1964,Wang2014a} has led to the development of several successful photoabsorber materials for thin-film solar cells.
For instance, the I-III-VI$_2$ chalcopyrite Cu(In,Ga)Se$_2$ (CIGS) is one of the most mature thin-film absorbers with record power conversion efficiency above 23\%.~\cite{Green2020} Branching out even further from the II-VI template, efficiencies above 12\% have been demonstrated by the I$_2$-II-IV-VI$_4$ kesterite Cu$_2$ZnSn(S,Se)$_4$ (CZTS).~\cite{Yan2018b} CZTS has a more favorable mix of earth-abundant and non-toxic elements than CIGS but is still limited by tail states and deep defects.~\cite{Hages2017,Rey2018}
With a growing number of absorbers materials demonstrating conversion efficiencies well above 20\% in single-junction solar cells,~\cite{Green2020} further progress in solar energy conversion is likely associated with the development of high-efficiency tandem cells combining a narrow- and a wide band gap absorber.
While there are obvious candidates to the role of narrow band gap absorber (e.g. silicon and CIGS), the ideal wide band gap absorber has, arguably, not been found yet. Among the few existing high-efficiency wide band gap absorbers, III-V semiconductors are not cost effective for large-scale applications, and it is still unclear whether the highly reactive halide perovskite semiconductors can be protected in the long term against degradation.

In the context of tandem solar cells, replacing Zn with an alkaline earth metal (Sr or Ba) in CZTS is \red{of particular interest}. First of all, band gaps in Cu$_2$BaSnS$_4$ (CBTS) and Cu$_2$SrSnS$_4$ (CSTS) are wider than in CZTS, and they can be tuned across the whole optimal range for a tandem cell top absorber by Se alloying.~\cite{Shin2017a} Additionally, the tail states that are predominant in CZTS are significantly mitigated in both CBTS and CSTS, as judged by the abruptness of their absorption onset, by the negligible absorption-emission Stokes shift, and by the outcome of pump-probe experiments.~\cite{Crovetto2019a,Crovetto2019b,Ghadiri2018} Furthermore, a few theoretical studies have found relatively low carrier effective masses and a favorable defect chemistry in both CBTS and CSTS.~\cite{Hong2016,Zhu2017,Pandey2018} Finally, CBTS and CSTS rely on earth-abundant elements without obvious toxicity concerns.~\cite{Vesborg2012} Despite these compelling features, experimental work on CBTS and CSTS is still in its infancy. Single-junction CBTS solar cells have reached 2.0\% efficiency with a pure sulfide absorber~\cite{Ge2017} and 5.2\% with Se alloying.~\cite{Shin2017a} Only two papers reporting CSTS solar cells are known,~\cite{Crovetto2019a,Xiao2020} with 0.6\% as the highest single-junction cell efficiency.~\cite{Crovetto2019a}

In this joint experimental and theoretical work, we characterize CBTS and CSTS films by a variety of spectroscopic techniques. When possible, experimental spectra are compared to simulated spectra based on the results of first-principles calculations.
To ensure consistency, films of the two materials are grown using the same method, based on sulfurization of reactively sputtered oxide precursors. We conclude that CBTS and CSTS are remarkably similar in their structural, vibrational, dielectric, and optical properties. 

Room-temperature photoluminescence (PL) spectra of CBTS and CSTS seem to indicate high-quality materials with low band tailing. However, the spectra change dramatically at lower temperatures, with clear indications of radiative recombination transitions involving both shallow defects and deep defects. Even though CBTS and CSTS are p-type semiconductors in the bulk, downward band bending is observed at their surface. In CBTS, this band bending appears to be strong enough to induce n-type conductivity at the surface, which can be a beneficial effect for enhancing carrier separation and keeping the main recombination path away from the heterointerface. Compared to CIGS and CZTS, the conduction bands of CBTS and CSTS lie at a much higher energy on an absolute scale. This low electron affinity implies that the CdS/ZnO electron contact traditionally used in many chalcogenide solar cells is likely not optimal for these absorbers.

\section{Experimental details}
CBTS and CSTS films were grown by sulfurization of oxide precursor films deposited by reactive sputtering. Details of the growth process are available in previous publications.~\cite{Crovetto2019a,Crovetto2019b} 1~$\mu$m-thick CBTS or CSTS films deposited on Mo-coated soda lime glass (SLG) were used for all characterization with the exception of ellipsometry measurements. To avoid modeling complications associated with large surface roughness and multilayer structures,~\cite{Crovetto2016,Crovetto2015} ellipsometry measurements were performed on 150-200~nm-thick films deposited on crystalline silicon. Structural characterization by x-ray diffraction (XRD) and compositional characterization by energy-dispersive x-ray spectroscopy (EDX) was carried out in previous studies.~\cite{Crovetto2019a,Crovetto2019b}

The vibrational Raman spectra of the two compounds were measured with a Thermo Scientific DXR Raman microscope in the backscattering configuration using a 633~nm or 780~nm laser and a 25~$\mu$m pinhole before the spectrometer. A 10X objective was used in conjunction with 1~mW laser power, giving an excitation density of approximately $\sim\SI{10}{W \per mm^2}$. We verified that the position and FWHM of the main Raman peak at this excitation level were unchanged with respect to the case of a lower laser power.
The optical dielectric functions of CBTS and CSTS were determined by spectroscopic ellipsometry in the near ultraviolet to near infrared range using a J.A. Woollam M-2000 rotating compensator ellipsometer. The imaginary part ($\varepsilon_2$) of the unknown dielectric function of the absorbers was fitted to a b-spline function with 8 nodes in the 1.8-2.2 eV region to capture the sharp feature in the absorption coefficient just above the band gap, and 5 nodes/eV elsewhere. Kramers-Kronig integration was then performed to derive the real part of the dielectric function ($\varepsilon_1$) minus the high-frequency dielectric constant ($\varepsilon_\infty$), which was also determined by least-squares fitting. Refractive index $n$, extinction coefficient $\kappa$, and absorption coefficient $\alpha$ were then derived using standard optical relations. The ellipsometry data analysis method has been previously described in detail.~\cite{Crovetto2016,Crovetto2018a} To estimate of the band gap and exciton binding energy in the presence of excitonic absorption, ellipsometry spectra between 1.8~eV and 2.2~eV were also fitted with an Elliot function~\cite{Elliott1957} with four fitting parameters (amplitude, band gap, exciton binding energy, and broadening parameter) and with two additional Lorentzian oscillators to model absorption features at photon energies slightly below and above the main absorption onset. For comparison, the same fitting procedure was performed on the near-band gap portion of unpolarized transmission spectra (also measured with the M-2000 ellipsometer) after the film thickness was determined by fitting the full transmission spectrum and kept fixed thereafter.

Low-temperature photoluminescence (PL) measurements and room-temperature microphotoluminescence mapping were performed with a customized scanning microscopy setup based on a Nikon Eclipse Ti-U inverted microscope and a continuous wave (CW) 523~nm laser. Using a beam splitter, laser light was focused on the sample by an objective lens and PL emission was collected by the same objective. Then, PL emission was filtered by a 550 nm long pass filter and directed to a spectrometer (Shamrock 303i, Andor) equipped with an electronically cooled CCD detector through a 250 $\mu$m input slit. For room-temperature microphotoluminescence mapping, the sample was scanned using a X-Y stepper motor-driven stage and a 50X objective, resulting in a spot size of $\sim\SI{1.5}{\micro\metre}$ and an excitation density of $\sim\SI{30}{W \per mm^2}$. For single-point low-temperature PL measurements, the sample was placed inside a temperature controlled stage (HFS600, Linkam Scientific Instruments) and a 10X objective was used, resulting in a spot size of $\sim\SI{11}{\micro\metre}$ and an excitation density of $\sim\SI{400}{mW \per mm^2}$.

Large-area PL spectra were used to compare the typical room-temperature PL emission of CBTS, CSTS, and CZTS, and correlate it to their external quantum efficiency (EQE). These large-area spectra were measured with an Accent RPM2000 system using a 405~nm CW excitation laser with a spot size of 1 mm$^2$ and excitation density of 1 W/mm$^2$. EQE was measured on completed CBTS, CSTS, and CZTS solar cells~\cite{Cazzaniga2017,Crovetto2019a,Crovetto2019b} using a PV Measurements QEXL setup calibrated with a reference Si photodiode.
Surface composition, core level positions, work function, and the position of the valence band maximum with respect to the Fermi level were determined by x-ray photoemission spectroscopy (XPS) using a Thermo Scientific K-Alpha instrument with a monochromatized Al K$_\mathrm{\alpha}$ x-ray source at 1486.68 eV at a base pressure below $5 \times 10^{-9}$~mbar. An Ar$^+$ ion beam at different beam energies was used for removing surface contamination. Elemental composition was determined using the following core levels: Cu 2p$_{3/2}$, Ba 3d$_{5/2}$, Sr 3d, Sn 3d$_{5/2}$, S 2p, O 1s, and Na 1s. The core level peaks were fitted with a single Voigt function and a Shirley background (Cu, Ba, Sr, Sn, S) or a linear background for the low-intensity peaks (O, Na) The films were analyzed immediately after sulfurization with sample transfer performed in an Ar-filled transfer box with minimal air exposure. Work function and valence band maximum (VBM) were measured by linear extrapolation of the photoemission threshold on the low kinetic energy edge and low binding energy edge of the spectrum, respectively. For the work function measurement, a voltage of $-30$~V was applied to the sample in order to shift the whole spectrum to higher kinetic energies and thus deconvolve the work functions of the sample and of the detector. The work function energy scale was calibrated with the known work function (5.1~eV) of an ion beam-cleaned strip of Au foil in contact with the sample. The energy scale used for determining the positions of core levels and VBM with respect to the Fermi level was calibrated with the Fermi edge of the same Au foil.

\section{Computational details}

The calculations started from the experimentally determined structures of CSTS~\cite{Teske1976} and CBTS~\cite{Teske1976a} as available in the Inorganic Crystal Structure Database (ICSD). Convergence tests regarding plane wave energy cut-off and $k$ point mesh revealed a combination of 400 eV energy cut-off and $5\times5\times3$ $k$ point mesh is sufficient to converge the total energy to \SI{1}{\milli\electronvolt} per atom. 
The geometry optimization was carried out using the PBEsol functional,\cite{Perdew2008} which is an non-empirical PBE functional revised for solids and believed to yield better predictions for equilibrium properties such as lattice parameters. The energy cut-off was increased to \SI{520}{\electronvolt} during geometry optimization to avoid Pullay stress raised by the incompleteness of the plane wave basis set. The structures were deemed fully relaxed when force on every atom fell below \SI{1E-4}{\electronvolt\per\angstrom}. Optical absorption was calculated using the HSE06 functional and a $7\times7\times3$ $k$ point mesh.\cite{Krukau2006}

Phonon dispersion calculations (Fig.~S1, Supporting Information) were carried out using finite atomic displacements of a 96-atom supercell of relaxed \ce{CSTS} and \ce{CBTS} generated by the Phonopy code,\cite{Togo2015} from which the force constants were obtained and used in Raman spectrum calculations.~\cite{Fleck2020}
Two displaced structures for each Raman active mode were generated using the Phonopy-Spectroscopy code,\cite{Skelton2017} and for each displaced structure the dielectric constant was evaluated using density functional perturbation theory (DFPT). 
Due to a discrepancy between experimental and calculated Raman spectra, we tested a number of alternative functionals besides PBEsol for structural relaxation. These included LDA and standard PBE, with and without an on-site Hubbard U correction\cite{Dudarev1998} of \SI{5.2}{\electronvolt} for copper.\cite{Kehoe2016} The resulting lattice constants are shown in Table~S2, Supporting Information.

\begin{figure}[t!]
\centering%
\includegraphics[width=\columnwidth]{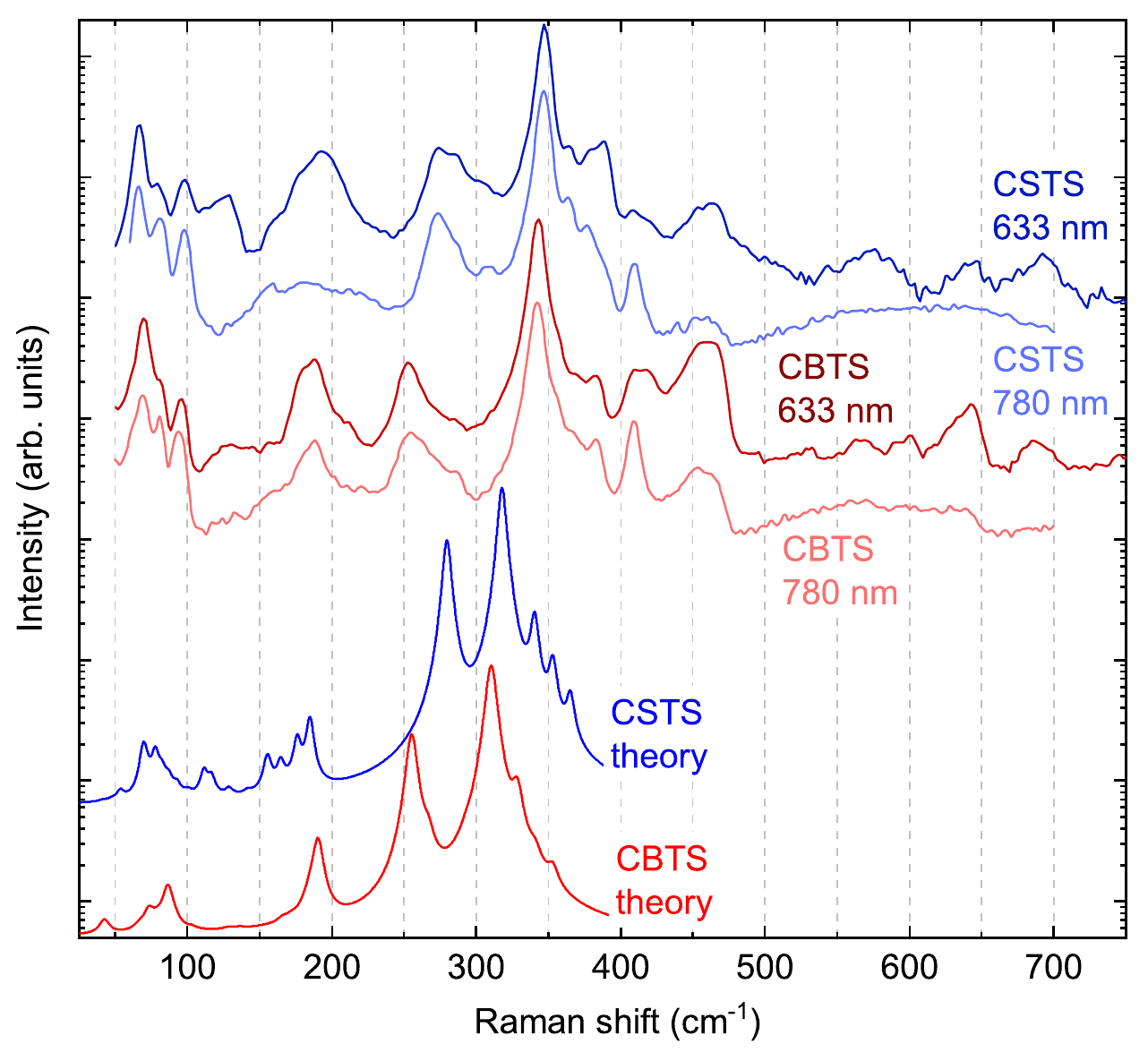}
\caption{Measured (top) and calculated (bottom) Raman spectra of CSTS and CBTS. Two sets of measured spectra are shown, differing in the excitation wavelength (633~nm and 780~nm). The position and symmetry type of each peak are shown in Tables~S3,S4, Supporting Information.}
\label{fig:raman}
\end{figure}

\section{Results and discussion}
\subsection{Raman spectroscopy}
Both CSTS and CBTS belong to space group $P3_1$. With reference to the character table, all optical modes are Raman active. For a 24-atom unit cell, there are then 23 A modes and 23 doubly degenerate E modes, giving rise to $23\times3 = 69$ modes, of which 46 have distinct vibrational frequencies. Computed and experimental Raman spectra at two excitation wavelengths are shown in Fig.~\ref{fig:raman}. The positions of the experimental and computed peaks are summarized in Tables~S3,S4, Supporting Information. The computed Raman spectra are in good agreement with the experimental spectra up to $\sim$300~cm$^{-1}$ Raman shift. At higher wavenumbers, the computed spectra are red-shifted by roughly 30~cm$^{-1}$ with respect to the experimental spectra (Fig.~\ref{fig:raman}). This discrepancy is reproduced with a range of exchange correlation functionals and with on-site Hubbard U correction, indicating that errors in the computed lattice constants are not responsible for the red shift. A similar discrepancy was observed between the experimental and computed Raman spectra of CZTS~\cite{Skelton2015,Ramkumar2016} and was tentatively related to the overestimated polarizability of S with gradient-corrected exchange-correlation functionals. Despite the red shift above $\sim$300~cm$^{-1}$, several relative trends in that spectral region are still captured by the calculation. For example, the main CSTS peak (347~cm$^{-1}$) occurs at a slightly higher wavenumber than the main CBTS peak (343~cm$^{-1}$) and has stronger satellite peaks on the high wavenumber side. Note also that excitation at 633~nm (1.96~eV) is nearly resonant with the band gaps of CBTS and CSTS. Accordingly, the intensity of the peaks related to E vibrational modes increases with respect to the case of sub-band gap excitation at 780~nm, due to their polar character.~\cite{Dimitrievska2014a}  Although all 69 first-order Raman modes are predicted to be below 400~cm$^{-1}$ in both materials, additional peaks are clearly observed at higher wavenumbers in the experimental spectra. The peaks above 450~cm$^{-1}$ are tentatively attributed to second-order Raman peaks of CBTS and CSTS, since they tend to increase in intensity under resonant excitation at 633~nm. Contrarily to all other peaks in the spectra, the intensity of the peak at 409-410~cm$^{-1}$ decreases using 633~nm excitation, so this peak is possibly related to a secondary phase. Although it is in principle compatible with MoS$_2$,~\cite{Fernandes2011} its intensity is over one order of magnitude higher than the MoS$_2$ peak measured on a bare MoS$_2$ film, so assignment to MoS$_2$ is excluded. A peak at 413~cm$^{-1}$ was previously identified in CBTS deposited on glass~\cite{Ge2017} but it had a lower relative intensity at the same excitation wavelength. Hence we speculate that the 409-410~cm$^{-1}$ peaks may be related to Sr$_2$SnS$_4$ and Ba$_2$SnS$_4$ secondary phases observed near the Mo back contact in our films.~\cite{Crovetto2019a,Crovetto2019b}

\begin{figure}[t!]
\centering%
\includegraphics[width=\columnwidth]{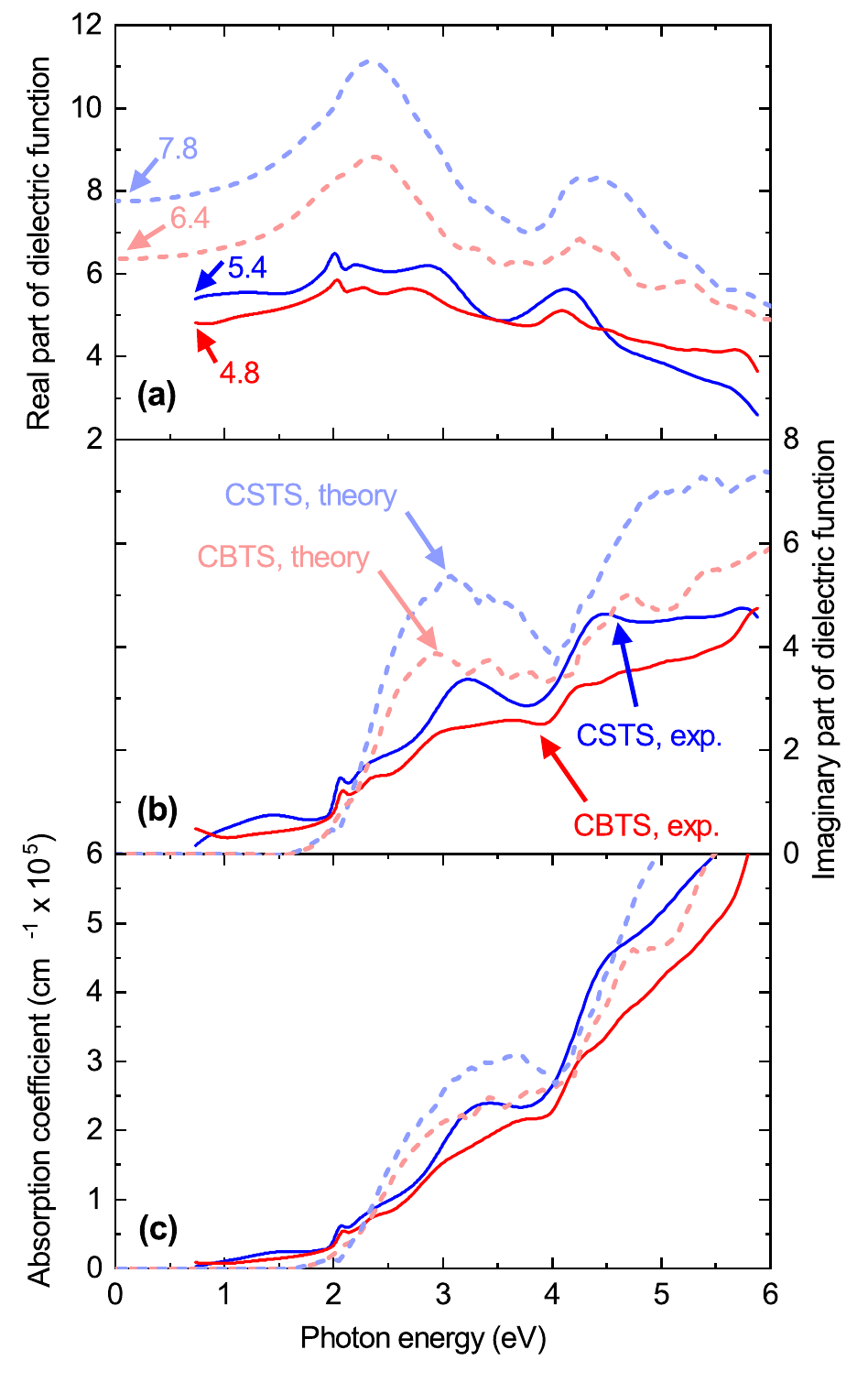}
\caption{Complex dielectric function (a,b) and absorption coefficient (c) of CSTS and CBTS. Experimental spectra (solid lines) are determined by ellipsometry using a b-spline function. Simulated spectra (dashed lines) are calculated with the HSE approach. Values of $\varepsilon_\infty$ are shown (note that experimental values are extrapolated). The refractive index and extinction coefficient of CSTS and CBTS are shown in Fig.~S2, Supporting Information.}
\label{fig:optical_functions}
\end{figure}

\subsection{Spectroscopic ellipsometry}
\begin{figure}[t!]
\centering%
\includegraphics[width=0.8\columnwidth]{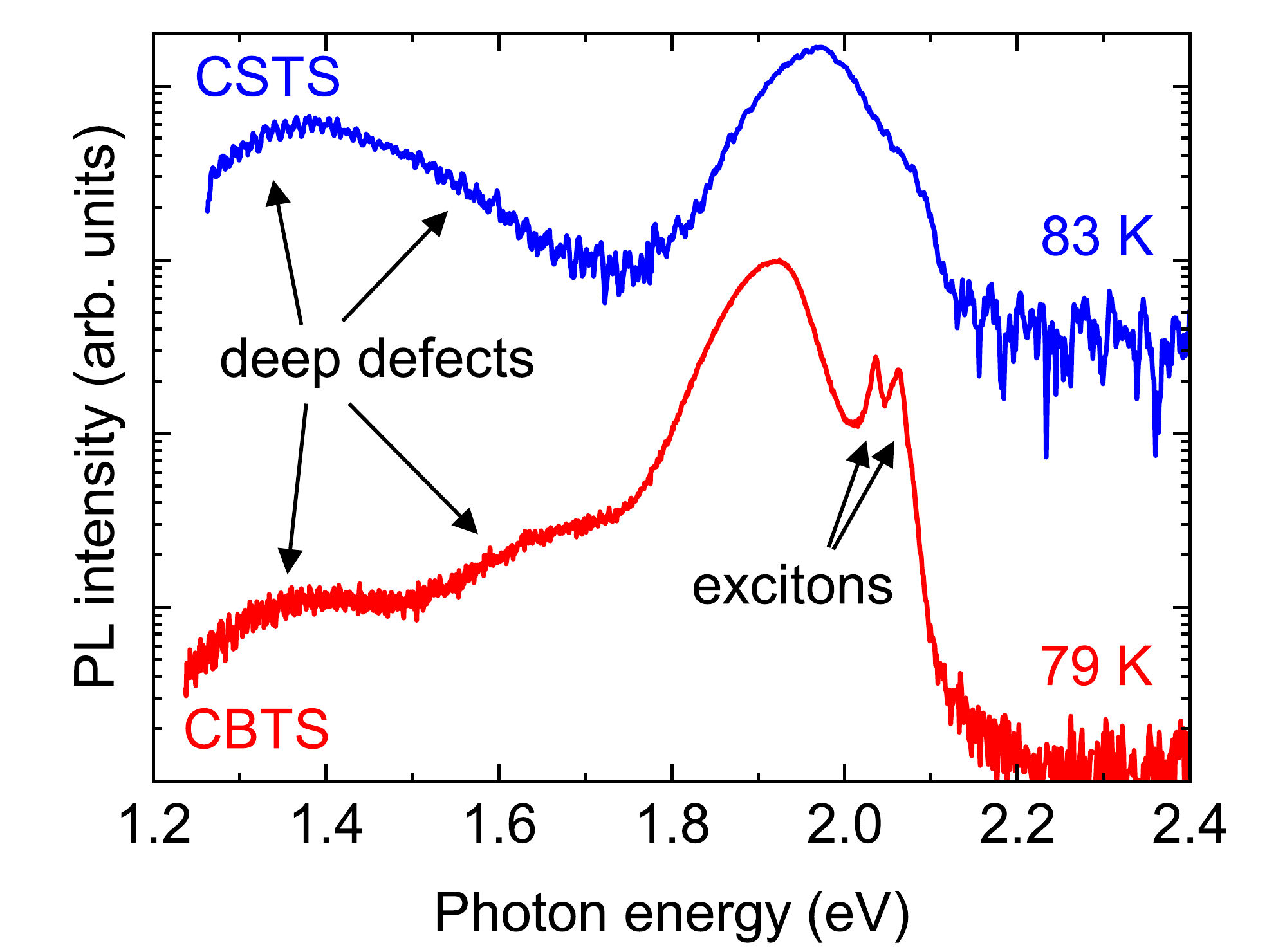}
\caption{PL spectra of CBTS and CSTS at 79~K and 83~K respectively. Note the low-intensity peaks at lower photon energies, associated with deep defects in the materials.}
\label{fig:PL_lowT}
\end{figure}

As shown in Fig.~\ref{fig:optical_functions}, CBTS and CSTS have rather similar dielectric function spectra. In the experimental spectra, a relatively sharp absorption onset \red{at around 2.0~eV} is followed by a dip, which is a sign of excitonic absorption.~\cite{Elliott1957} The observation of two narrow peaks in the PL spectra of CBTS at $\sim$80~K (Fig.~\ref{fig:PL_lowT}) reinforces this hypothesis, since such narrow peaks are typically due to excitonic transitions. As the HSE computational approach does not include effects related to excited states, the dip in $\varepsilon_2$ is not present in the calculated dielectric functions. The sharp absorption onset implies that photons with energy just above the band gap will be absorbed close to the front contact of the solar cell. As the absorption coefficient is around $5 \times 10^4$~cm$^{-1}$ just above the band gap, the majority of those long wavelength photons are generated within $1/\alpha \sim 200$~nm below the front contact. This shallow generation depth has a beneficial effect on carrier collection efficiency and on the minimum absorber thickness required for full sunlight absorption.

The large sub-band gap absorption in both materials is most likely an artifact due to a relatively rough film surface, as explained in detail in previous work.~\cite{Fujiwara2016} Although thinner films were employed for ellipsometry characterization to minimize surface roughness, we were not able to obtain a root-mean-square roughness below $\sim$8~nm with our synthesis method based on oxide precursors. The calculated values of $\varepsilon_\infty$ are 6.4 for CBTS and 7.8 for CSTS, which are somewhat larger than the experimental values of 4.8 and 5.4 respectively. Note, however, that the sub-band gap dielectric function artifacts discussed above are likely to introduce a significant error in the experimental values of $\varepsilon_\infty$. Experimental refractive indices in the transparent region below the band gap (2.2. for CBTS, 2.3 for CSTS) are also slightly lower than the calculated values (2.5 for CBTS, 2.8 for CSTS) as shown in Fig.~S2, Supporting Information.

\begin{figure}[t!]
\centering%
\includegraphics[width=\columnwidth]{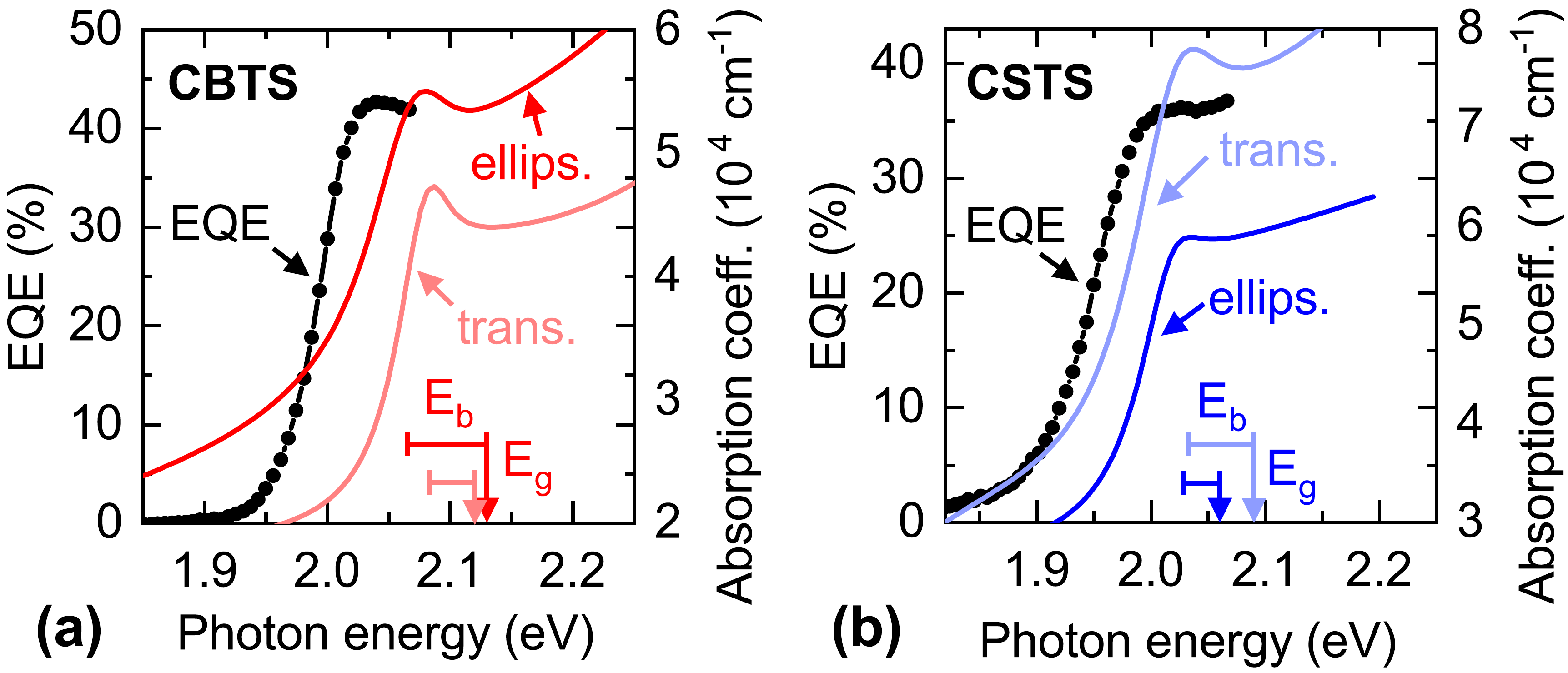}
\caption{Near-band gap absorption coefficients of CBTS (a) and CSTS (b) determined by fitting ellipsometry spectra or transmission spectra with the Elliott function,~\cite{Elliott1957} which includes the band gap $E_\mathrm{g}$ and exciton binding energy $E_\mathrm{b}$ as fitting parameters. The best-fit values of $E_\mathrm{g}$ and $E_\mathrm{b}$ from the two types of spectra are shown with the same color code as the absorption coefficients. The band gap is indicated with a vertical arrow, and the exciton binding energy is indicated with a horizontal line. The EQE of a CBTS solar cell and of a CSTS solar cells are also plotted in (a) and (b) respectively.}
\label{fig:elliott}
\end{figure}

The calculations predict that both $\varepsilon_1$ and $\varepsilon_2$ should overall be slightly larger in CSTS than in CBTS. This prediction is qualitatively reproduced by experiment, even though the measured dielectric functions have an overall lower magnitude than the calculated dielectric functions. Using the absorption coefficient measured by ellipsometry, Tauc plots for direct band gap materials yield a 2.00~eV band gap for CBTS and a 1.98~eV band gap for CSTS.~\cite{Crovetto2019a,Crovetto2019b} However, Tauc plots are not an appropriate method to determine band gaps from an excitonic absorption onset, since the latter does not reflect the density of states in the material. Here we estimate the band gap and exciton binding energy of CBTS and CSTS by fitting ellipsometry and optical transmission spectra near the band gap with an Elliott function.~\cite{Elliott1957} The resulting absorption coefficients are shown in Fig.~\ref{fig:elliott}. In CBTS, transmission (ellipsometry) data yield a band gap of 2.12~eV (2.13~eV) and an exciton binding energy of 37~meV (65~meV). In CSTS, transmission (ellipsometry) data yield a band gap of 2.09~eV (2.06~eV) and an exciton binding energy of 57~meV (32~meV). Thus, the band gaps of these compounds are likely to be about 100~meV larger than previously determined using Tauc plots. The error bar is rather large for two main reasons: (i) broadening of the excitonic absorption feature at room temperature, causing some degree of correlation between fitting parameters; (ii) the large apparent sub-band gap absorption, which has to be deconvolved from the absorption onset before fitting the spectra with the Elliott function. Taking the static dielectric constant and the direction-averaged electron- and hole effective masses from other computational works,~\cite{Zhu2017,Pandey2018,Crovetto2020c} the hydrogen model predicts exciton binding energies of 65~meV for CBTS and 62~meV for CSTS, in fair agreement with the measured values. The existence of appreciable excitonic absorption at room temperature in CBTS and CSTS is a unique feature of these compounds with respect to CZTS or most other kesterite-inspired materials, and it resembles the absorption features of some halide perovskites instead.~\cite{Ishihara1990,DInnocenzo2014} It can be explained by the lower dielectric constant, wider band gap, and higher hole effective masses in CBTS and CSTS with respect to CZTS.~\cite{Zhu2017,Pandey2018,Crovetto2020c}
The spectral onset of the photocurrent, based on external quantum efficiency (EQE) measurements of CBTS and CSTS solar cells, is plotted in Fig.~\ref{fig:elliott} together with the optically-detemined absorption coefficients. Interestingly, the photocurrent onset in both materials occurs at a lower photon energy than the band gaps determined using the Elliott function. The most likely explanation is that excitons dissociate at the contacts and contribute to the photocurrent as in, e.g., organic solar cells.~\cite{Green2014} The large cliff-like conduction band offset with the typical CdS heterojunction partner (see later) is expected to promote exciton dissociation.

\subsection{Photoluminescence spectroscopy}
\begin{figure}[t!]
\centering%
\includegraphics[width=\columnwidth]{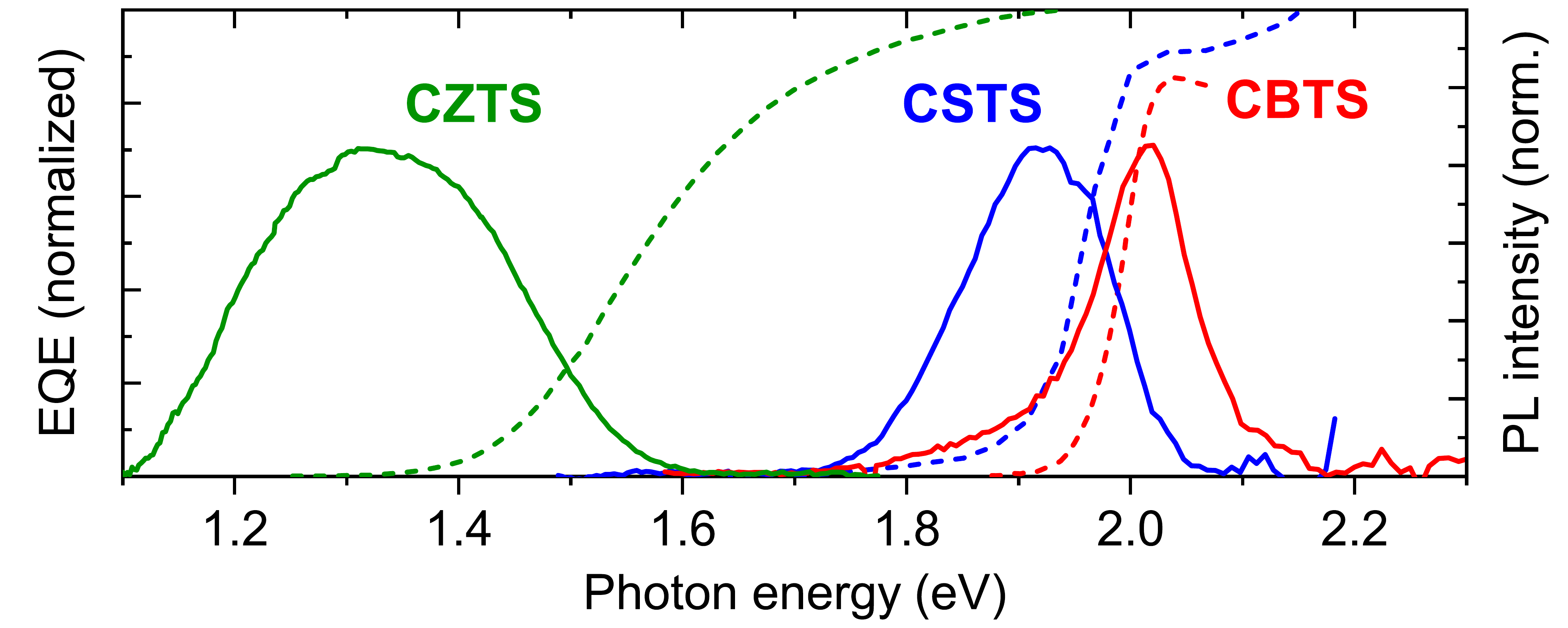}
\caption{Room-temperature PL spectra (solid lines) and EQE (dashed lines) of CZTS, CSTS, and CBTS films and solar cells. The width of the PL peak and its red shift with respect to the band gap increase in the order CBTS < CSTS < CZTS. All intensities are normalized.}
\label{fig:PL_EQE}
\end{figure}

As noted in previous work,~\cite{Crovetto2019a,Shin2017a} room-temperature PL spectra in CBTS and CSTS have certain compelling features. Both materials have a narrower peak and a smaller Stokes shift compared to CZTS (Fig.~\ref{fig:PL_EQE}). These features indicate that room-temperature emission in both materials may arise from band-to-band or exciton recombination, with negligible contributions from tail states. The abrupt onsets of optical absorption (Fig.~\ref{fig:optical_functions}(b)) and of the photocurrent (Fig.~\ref{fig:PL_EQE}) confirm this hypothesis. We argue that exciton recombination is the most likely origin of room-temperature PL for two reasons: (i) the optical absorption onset of CBTS and CSTS has a significant excitonic contribution even at room temperature, as discussed in the previous section; (ii) the position of the main PL peak as a function of temperature is not consistent with a transition from exciton recombination to band-to-band recombination with increasing temperature, as discussed in Ref.~\citenum{Crovetto2020c}.

\begin{figure}[t!]
\centering%
\includegraphics[width=\columnwidth]{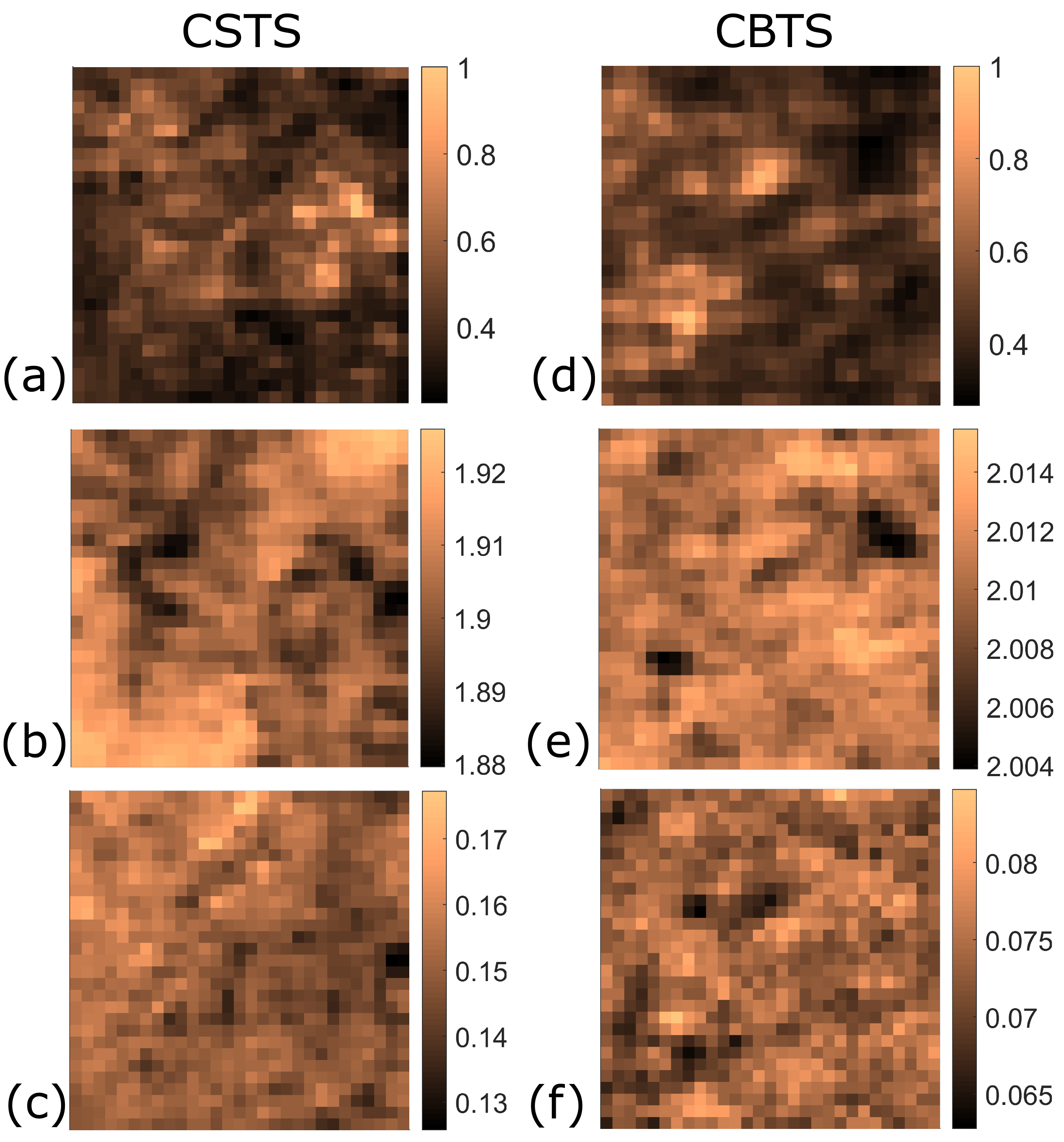}
\caption{PL maps of CSTS (a-c) and CBTS (d-f) at room temperature. (a) and (d) are maps of the normalized integrated PL intensity; (b) and (e) are maps of the PL peak position; (c) and (f) are maps of the FWHM of the PL peak. The size of each optical image and map is $\SI{14}{}$ x $\SI{14}{\micro\metre}$ with a step size of 0.5 $\mu$m and a spatial resolution of $\approx \SI{1.5}{\micro\metre}$.}
\label{fig:PL_map}
\end{figure}

Micro-PL maps over $14 \times 14~\mu$m regions (Fig.~\ref{fig:PL_map}(a-c)) reveal that the PL features of CBTS are fairly uniform at the microscale. With a spatial resolution of  $\sim 1.5~\mu$m, the standard deviation of PL intensity, peak position, and full width at half maximum (FWHM) is 23\%, 1.7~meV, and 3.2~meV respectively. Spatial PL inhomogeneity is more pronounced in CSTS (Fig.~\ref{fig:PL_map}(d-f)), with corresponding standard deviations of 25\%, 9.2~meV, and 6.6~meV for PL intensity, peak position, and FWHM.

At a temperature of $\sim$80~K, the PL characteristics of both compounds change significantly with respect to their room-temperature spectra (Fig.~\ref{fig:PL_lowT}). Temperature- and excitation-dependent PL of CBTS and CSTS are discussed in detail elsewhere.~\cite{Crovetto2020c}
Here we just focus on three qualitative features of the low-temperature spectra. First, both spectra are dominated by a broad peak that is significantly red-shifted with respect to the band gap (note that the band gaps of CBTS and CSTS increase with decreasing temperature).~\cite{Ge2017} This peak is related to shallow defect transitions which are quenched at room temperature. Second, the CBTS spectrum has two higher-energy peaks that are significantly narrower even than the room-temperature peak. They are related to excitonic transitions, probably a free exciton and a bound exciton peak.~\cite{Crovetto2020c} The CSTS spectrum does not exhibit such clear features, but its main peak has a high-energy shoulder, which is also attributed to exciton recombination.~\cite{Crovetto2020c}
Third -- and most importantly for optoelectronic applications -- two additional PL peaks are recognizable at around 1.4~eV and 1.6~eV photon energy in both CBTS and CSTS. Therefore, at least one radiatively active deep defect (up to 700~meV away from a band edge) must be present in both materials.
Deep defects are usually detrimental for the electronic quality of semiconductors, so we conclude that further progress in the photovoltaic efficiency of CBTS and CSTS is likely linked to a reduction in the density of these deep defects.

\subsection{X-ray photoemission spectroscopy}
\begin{figure}[t!]
\centering%
\includegraphics[width=\columnwidth]{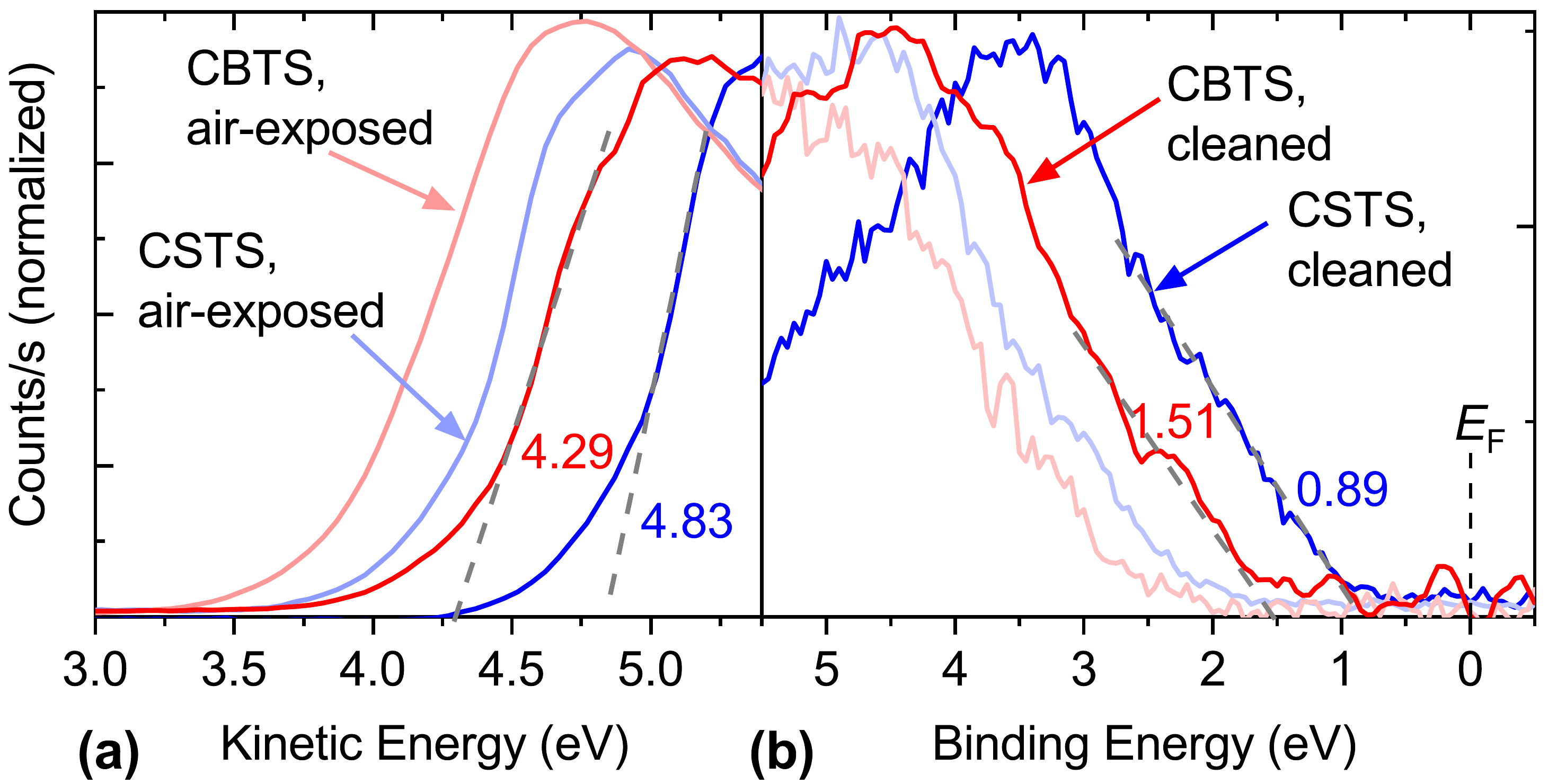}
\caption{Low kinetic energy onset (a) and low binding energy onset (b) of XPS spectra of CBTS (red) and CSTS (blue). The work function (a) and surface VBM (b) are extracted by linear extrapolation. Spectra are shown for the case of the air-exposed surface without ion-beam cleaning (pale colors) and the surface cleaned with a 100~eV ion beam (bright colors). Work function and VBM values after surface cleaning are indicated.}
\label{fig:xps}
\end{figure}

XPS compositional analysis of the as-sulfurized, air-exposed surfaces reveals about 10\% O and 3\% Na atomic composition (Fig.~S3, Supporting Information). Na diffuses from the soda lime glass substrate, whereas O originates from a brief exposure to air during sample transfer and is absent in the bulk.~\cite{Crovetto2019a,Crovetto2019b} Most of the measured core level energies depend strongly on the parameters of the Ar$^+$ ion beam used to remove contaminants from the film surface before XPS analysis. Compared to the as-sulfurized surface, prolonged bombardment at 100~eV beam energy (Figs.~S3,S4, Supporting Information) only causes small changes to the overall composition, to the O and Na content, and to the core level positions and their full width at half maxima (FWHM). However, 100~eV bombardment causes a significant increase in the work function (Fig.~\ref{fig:xps}(a)) and an even more pronounced shift of the VBM closer to the Fermi level (Fig.~\ref{fig:xps}(b)). Based on these observations, we assume that this mild cleaning treatment removes most of the adsorbed species responsible for a surface dipole, but it is below the sputter threshold for removing the Na-containing metal oxide species at the surface.

At higher beam energies, O and Na at the surface are removed (Fig.~S3(c), Supporting Information) and most core levels shift by several hundreds meV (Fig.~S4, Supporting Information). However, these changes are accompanied by an increase in the peak FWHM and by the preferential loss of Sn and S (Fig.~S3(a,b), Supporting Information). As the Sn and S losses are roughly equal to each other at all ion beam energies, we attribute these losses to preferential sputtering of SnS, similarly to the case of CTZS.~\cite{Crovetto2018b} This phenomenon is not surprising, as SnS is the most volatile of the constituent binaries in CBTS and CSTS.~\cite{Rumble2019}
We choose to report the VBM and work functions measured after 20~min cleaning at 100~eV energy for the following reasons: (i) preferential loss of certain elements can change the charge distribution and the average oxidation state of each element, which can distort the core level positions and Fermi level; (ii) since the surface O and Na species are not removed by ion beam sputtering at 100~eV, they are also unlikely to be removed by subsequent processing and thus are an integral part of the final device; (iii) the surface O content is only a small fraction of the S content and is unlikely to introduce a significant distortion with respect to a pristine surface without air exposure.

\begin{figure}[t!]
\centering%
\includegraphics[width=\columnwidth]{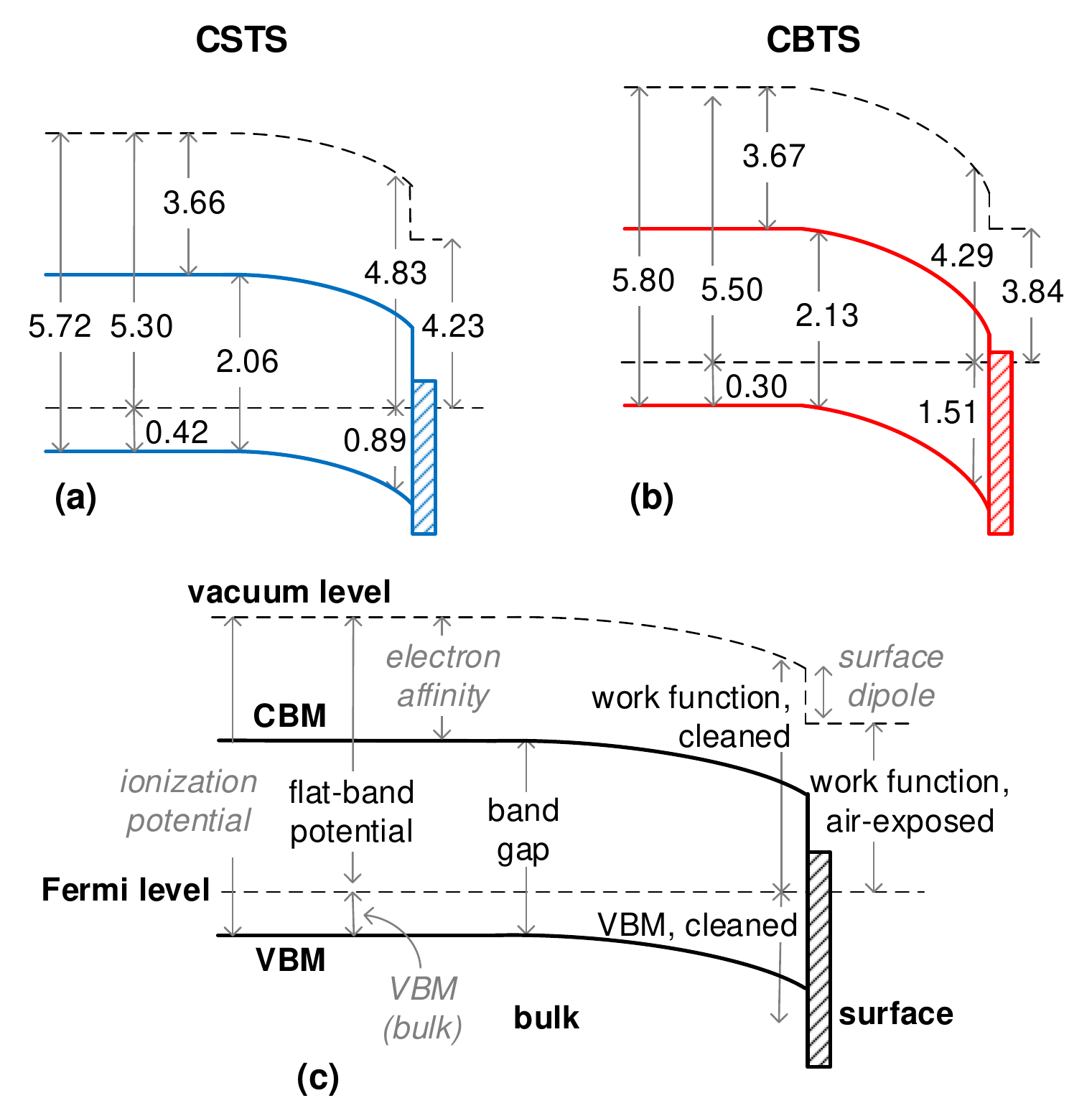}
\caption{(a,b) Experimental band diagrams of the bulk and surface of CSTS (a) and CBTS (b). (c) Explanation of the quantities shown in the band diagram. The measured quantities (in regular font) are used to derive additional material properties (in italics). Work functions and VBM position at the surface are determined by XPS, the flat-band potential is determined by electrochemical Mott-Schottky analysis,~\cite{Ge2017,Crovetto2019a} and the band gap is determined by fitting near-band-gap ellipsometry spectra with an Elliott function (Fig.~\ref{fig:elliott}).}
\label{fig:surface_band_diagram}
\end{figure}

By linear extrapolation of the photoemission onsets, we extract work function values of 4.29~eV (CBTS) and 4.83~eV (CSTS). In a similar fashion, the VBM is 1.51~eV below the Fermi level in CBTS and 0.89~eV below the Fermi level in CSTS (Fig.~\ref{fig:surface_band_diagram}). Such low work functions and high Fermi levels with respect to the VBM are not expected for p-type absorbers with relatively wide band gaps. In fact, the surface Fermi level is close to mid gap in CSTS, and it is closer to the conduction band than to the valence band in CBTS, implying an inverted (n-type) surface as opposed to a p-type bulk in CBTS. The \textit{surface-sensitive} quantities obtained by XPS (analysis depth of a few nm) can be compared to the \textit{bulk} flat-band potential obtained by capacitance-based electrochemical measurements~\cite{Crovetto2019a,Ge2017} to draw more complete band diagrams for CBTS and CSTS. From measured flat-band potentials, the position of the bulk Fermi level with respect to the vacuum level can be calculated as 5.5~eV for CBTS~\cite{Ge2017} and 5.3~eV for CSTS.~\cite{Crovetto2019a} Taking the band gaps of CBTS and CSTS as 2.13~eV and 2.06~eV respectively (Fig.~\ref{fig:elliott}) the band diagrams shown in Fig.~\ref{fig:surface_band_diagram} can be plotted. The low work function measured by XPS can be explained by downward band bending at the surface. The band diagram can be reconciled with the observation of p-type bulk doping in both compounds by combining the flat-band potential and the XPS data to derive the VBM position in the bulk, which is indeed close to the Fermi level in both compounds (0.30~eV below the Fermi level in CBTS, 0.42~eV in CSTS). By summing the work function and VBM position measured by XPS and subtracting the band gap, nearly equal electron affinities are derived for CBTS (3.67~eV) and CSTS (3.66~eV). These values are consistent with a simple empirical model of band positions based on the geometric mean of the Mulliken electronegativities of the constituent elements,~\cite{Butler1978} which predicts electron affinities of 3.65~eV in CBTS and 3.72~eV in CSTS. Since the electron affinities of CIGS and CZTS are instead in the 4.2-4.6~eV range,~\cite{Repins2011a} the CdS heterojuction partner typically used for those absorbers is probably not optimal for transporting electrons from CBTS and CSTS absorbers. A very large cliff-type conduction band alignment is expected for the CBTS/CdS and CSTS/CdS interfaces, which will ultimately limit the open circuit voltage in the corresponding solar cells.~\cite{Crovetto2017b} Conversely, type inversion at the absorber surface is generally considered as a beneficial effect for enhancing carrier separation and for keeping the main recombination path away from the heterointerface.~\cite{Shafarman2010}

\section{Conclusion}
Due to the chemical similarity of Ba and Sr, CBTS and CSTS have remarkably similar structural parameters and electronic structure.~\cite{Hong2016,Crovetto2019a,Crovetto2019b} Hence, only subtle differences were observed in their vibrational spectra, band gaps, dielectric properties, optical absorption, and absolute band positions by the spectroscopic techniques employed in this work. However, the room-temperature PL peak in CBTS is narrower, less Stokes-shifted, and has less microscale inhomogeneity than in CSTS. At lower temperatures the PL spectral features change dramatically in both compounds, as new peaks related to transitions involving shallow and deep defects appear. Optical absorption at room temperature is exciton-enhanced in both CBTS and CSTS, so Tauc plots are not an appropriate method to determine the band gaps of these materials. Deconvolution of excitonic effects using an Elliott function yields band gaps that are about 100~meV higher than previous estimates based on Tauc plots. 
CBTS and CSTS are p-type semiconductors in the bulk, but downward band bending is observed at their surfaces. According to XPS, band bending in CBTS is large enough to induce n-type surface conductivity, which can be a beneficial effect for enhancing carrier separation and for keeping the main recombination path away from the heterointerface. At the device level, it is important to recognize that the conduction bands of CBTS and CSTS lie at a significantly higher energy than in CIGS and CZTS. Hence, the CdS/ZnO electron contact often used in chalcogenide solar cells is likely not optimal for these absorbers and low-electron affinity contact materials, \red{such as (Zn,Mg)O}, should investigated instead.
Although there are clear indications that tail states in CSTS and (especially) CBTS are less severe than in CZTS, there are certainly other factors that can hinder success of a new photovoltaic absorber. Based on the results of the present study, we identify two important priorities for future research in these materials: (i) better understanding of deep defects and development of growth methods which can suppress their formation; (ii) investigation of alternative heterojunction partners and electron contacts with a low electron affinity.

%
%
%

\section*{Acknowledgements}
This project has received funding from VILLUM Fonden (grant no. 9455) and from the European Union’s Horizon 2020 research and innovation programme under the Marie Sklodowska-Curie grant agreement No 840751. The Center for Nanostructured Graphene is sponsored by the Danish National Research Foundation (Project No. DNRF103). The PL imaging setup has been partly funded by the IDUN Center of Excellence funded by the Danish National Research Foundation (project no. DNRF122) and VILLUM Fonden (grant no. 9301). Calculations were performed on the Archer HPC system, via the UK Materials Chemistry Consortium (EPSRC P/L000202), and the UCL Legion and Grace facilities. We are also grateful for resources from the UK Materials and Molecular Modelling Hub (EPSRC EP/P020194/1).

\section{References}
%

\newpage
\clearpage


\onecolumngrid

\section*{SUPPORTING INFORMATION}

\vspace{2cm}

\renewcommand{\thefigure}{S1}
\begin{figure}[h!]
\centering%
\includegraphics[width=0.4\textwidth]{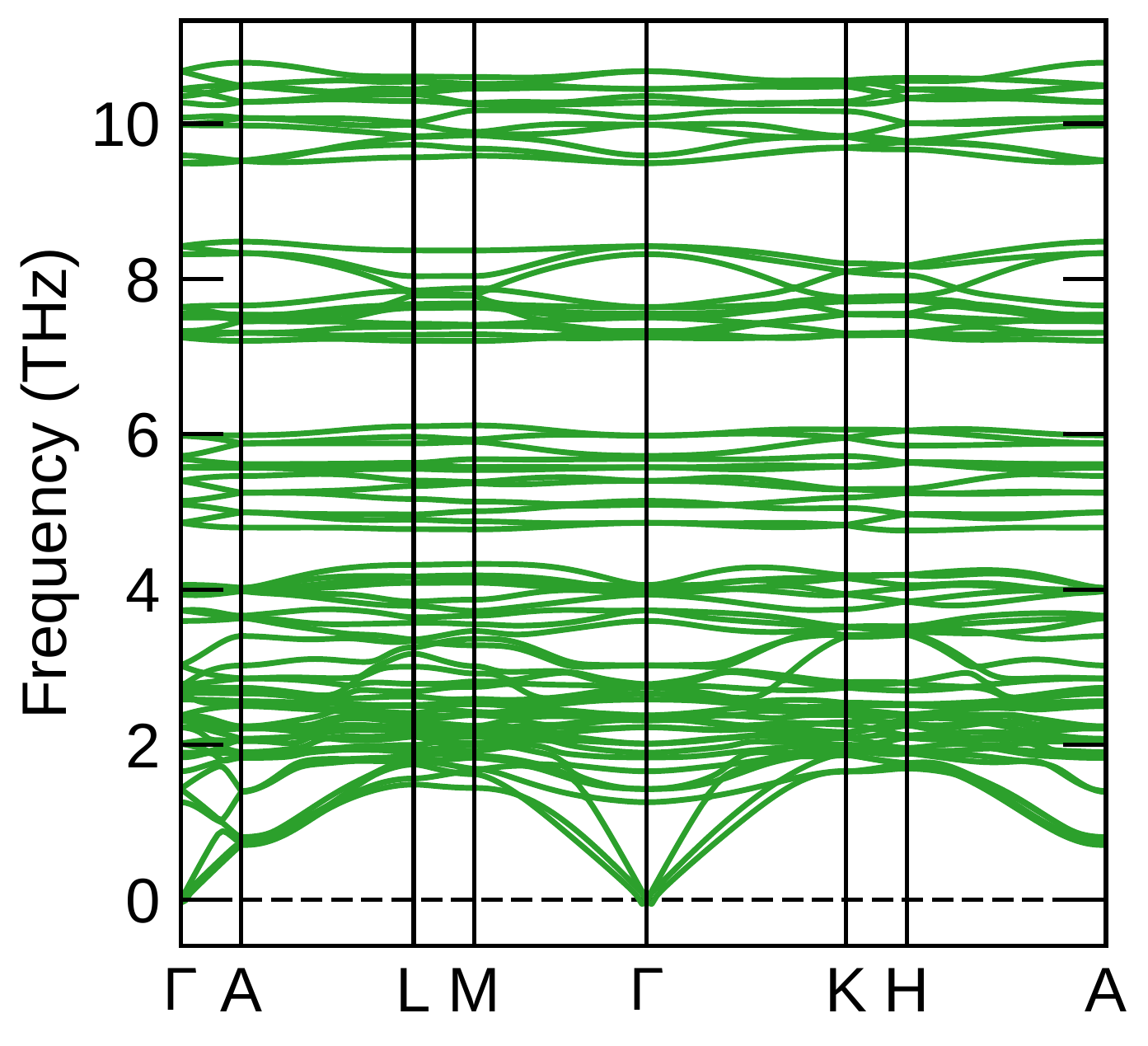}
\caption{Phonon band structure calculated with the PBEsol functional.}
\label{fig:phonons}
\end{figure}

\vspace{1cm}

\renewcommand{\thetable}{S1}
\begin{table}[h!]
\centering
\begin{tabular}{l c c c c}
\hline
PBEsol & $a$ (\AA) & $c$ (\AA) & $\alpha$ ($^\circ$) & $\gamma$ ($^\circ$) \\
\hline
\ce{CSTS} & 6.249 (-0.65\%) & 15.364 (-1.34\%) & 90.00 (0.00\%) & 120.00 (0.00\%) \\
\ce{CBTS} & 6.337 (-0.46\%) & 15.580 (-1.63\%)\ & 90.00 (0.00\%) & 120.00 (0.00\%) \\
\hline
\end{tabular}
\caption{PBEsol relaxed lattice parameters of \ce{CSTS} and \ce{CBTS} unit cell, percentage difference from experiments given in brackets\cite{TeskeA,TeskeB}. CBTS and CSTS crystallise into the same $P3_1$ space group.}
\label{geo_relax}
\end{table}

\vspace{1cm}

\renewcommand{\thetable}{S2}
\begin{table}[h!]
\centering
\begin{tabular}{l c c c c}
\hline
functionals& $a$ (\AA) & $c$ (\AA) & $\alpha$ ($^\circ$) & $\gamma$ ($^\circ$) \\
\hline
LDA & 6.273 (-1.48\%) & 15.408 (-2.68\%)\ & 90.00 (0.00\%) & 120.00 (0.00\%) \\
LDA+U & 6.269 (-1.55\%) & 15.425 (-2.57\%)\ & 90.00 (0.00\%) & 120.00 (0.00\%) \\
PBE & 6.450 (+1.31\%) & 15.854 (+0.14\%) & 90.00 (0.00\%) & 120.00 (0.00\%) \\
PBE+U & 6.448 (+1.28\%) & 15.886 (+0.33\%)\ & 90.00 (0.00\%) & 120.00 (0.00\%) \\
PBEsol & 6.337 (-0.46\%) & 15.580 (-1.63\%)\ & 90.00 (0.00\%) & 120.00 (0.00\%) \\
PBEsol+U & 6.333 (-0.54\%) & 15.6024 (-1.46\%)\ & 90.00 (0.00\%) & 120.00 (0.00\%) \\

\hline
\end{tabular}
\caption{Lattice parameters of \ce{CBTS} unit cell relaxed using different functionals, percentage difference from experiments given in brackets\cite{TeskeA}.}
\label{geo_relax}
\end{table}

\renewcommand{\thetable}{S3}
\begin{table}[htb]
\centering
\begin{tabular}{c c c c}
\hline
Raman shift (experiment)  & Raman shift (theory) & Intensity (theory) & Symmetry (theory) \\
cm$^{-1}$  & cm$^{-1}$ & arb. units &  \\
\hline
69		& 42.5 	& 2096 		 & A  \\
		& 47.6 	& 1		     & E \\
		& 54.5 	& 0.3		 	 & A  \\
		& 59.2 	& 1 			 & A  \\
    	& 63.0 	& 17			 & A  \\
		& 66.6 	& 670		 & E \\
81		& 73.0 	& 2441		 & E  \\
		& 74.5 	& 947		 & E  \\	
	  	& 79.1 	& 763		 & A  \\
		& 80.5 	& 407    	 & E \\
95		& 86.4 	& 9110		 & A  \\
		& 88.8 	& 672	     & E  \\		
	    & 89.8 	& 672		 & A  \\
		& 91.5	& 454    	 & E \\
		& 93.4 	& 0.1		     & A  \\
		& 103.1 & 313        & E  \\		
        & 125.2 & 0.4		 & A  \\
		& 127.6	& 233  	     & E \\
		& 135.7 & 122		 & A  \\
	    & 136.1 & 107        & E  \\
	    & 140.1 & 0.5		 & A  \\
		& 140.6	& 19	  	     & E \\
		& 166.1 & 687        & E  \\
		& 172.0 & 515        & A  \\		
        & 174.8	& 0.1		   	 & A  \\
		& 180.9	& 713     	 & E \\
182	& 185.2 & 3026	     & E  \\
		& 189.0 & 14		     & A  \\
188   & 190.4	& 32020   	 & A  \\
		& 198.3	& 102	  	 & E \\
		& 252.7 & 4137 	     & E  \\
		& 255.0 & 6903       & A  \\		
255   & 255.3 & 279877	 & A  \\
		& 265.1	& 1870	  	 & E \\
		& 266.2 & 66449     & E  \\
285	& 267.2 & 10648     & A  \\		
		& 292.9 & 4371	     & E  \\
		& 293.0 & 4		     & A  \\
  	    & 306.8	& 11204  	 & E  \\
343	& 310.4	& 1114066	 & A \\		
355	& 328.3 & 72594	 & E  \\
		& 331.3 & 11		     & A  \\
367   & 340.1	& 6313	 	 & A  \\
		& 341.1	& 8	    	 & A \\		
	    & 342.2	& 2682     	 & E  \\
383	& 353.0	& 7189    	 & E \\
409   &         &     	         &   \\
453   &         &     	         &   \\
463   &         &     	         &   \\

\hline
\end{tabular}
\caption{CBTS Raman peak table}
\label{CBTS_raman_peaks}
\end{table}

\renewcommand{\thetable}{S4}
\begin{table}[htb]
\centering
\begin{tabular}{c c c c}
\hline
Raman shift (experiment)  & Raman shift (theory) & Intensity (theory) & Symmetry (theory) \\
cm$^{-1}$  & cm$^{-1}$ & arb. units &  \\
\hline
		& 41.6 	& 45 			 & A  \\
		& 47.3 	& 46 			 & E \\
		& 53.5 	& 402		 & A  \\
		& 58.7 	& 0.2 		 & A  \\
66		& 69.4 	& 3004		 & E  \\
		& 70.3 	& 558		 & E \\
		& 71.2 	& 603		 & A  \\
		& 73.2 	& 83		     & E  \\
81  	& 77.7 	& 2914		 & A  \\
		& 82.4 	& 730    	 & A \\
		& 82.5 	& 443		 & E  \\
		& 87.1 	& 854	     & E  \\
	    & 90.4 	& 16			 & A  \\
		& 92.9	& 544    	 & E \\
		& 93.4 	& 2		     & A  \\
		& 100.3 & 236        & E  \\
98    & 111.4 	& 1431		 & A  \\
		& 116.6	& 1001  	 & E \\
		& 120.8 & 17		     & A  \\
128	& 128.8 & 341        & E  \\
	    & 141.7 & 15		     & A  \\
		& 141.9	& 151	  	 & E \\
158	& 154.9 & 1536	     & E  \\
		& 156.2 & 341        & A  \\
        & 164.5	& 1185   	 & A  \\
		& 164.8	& 604     	 & E \\
		& 170.2 & 128	     & A  \\
179	& 176.0 & 4051       & E  \\
        & 184.8	& 3307   	 & A  \\
194	& 184.9	& 4145  	 & E \\
		& 273.2 & 2    	     & A  \\
		& 275.7 & 3651       & E  \\
273   & 279.7 & 298577   & A  \\
		& 281.1	& 600	  	 & E \\
285	& 283.9 & 4114       & E  \\
		& 285.6 & 1676       & A  \\
308	& 310.4 & 1325	     & E  \\
		& 317.3 & 2807      & A  \\
347   & 317.8	& 826363  	 & A  \\
		& 323.5	& 9492     	 & E \\
364	& 340.4 & 59354	 & E  \\
		& 341.1 & 580        & A  \\
377   & 352.9	& 15172  	 & A  \\
		& 353.1	& 7146     	 & E \\
	    & 355.2	& 26      	 & A  \\
		& 365.2	& 9773    	 & E \\
410   &         &     	         &   \\
464   &         &     	         &   \\

\hline
\end{tabular}
\caption{CSTS Raman peak table}
\label{CSTS_raman_peaks}
\end{table}

\renewcommand{\thefigure}{S2}
\begin{figure}[t!]
\centering%
\includegraphics[width=0.5\textwidth]{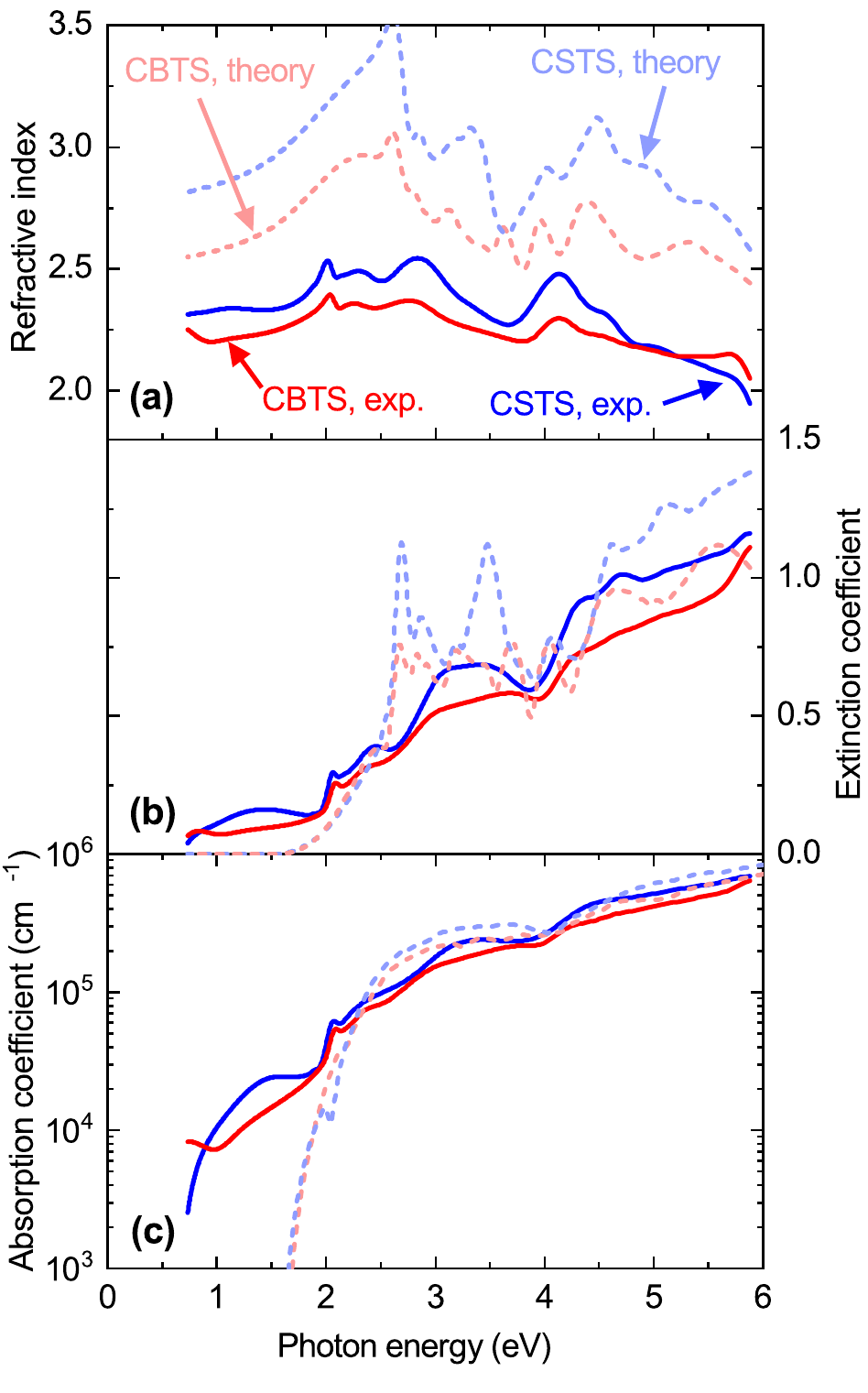}
\caption{Measured (solid lines) and calculated (dashed lines) optical properties of CBTS (red) and CSTS (blue). (a): Refractive index $n$; (b): extinction coefficient $\kappa$; (c) absorption coefficient $\alpha$ on a logarithmic scale.}
\label{fig:refractive_extinction}
\end{figure}

\renewcommand{\thefigure}{S3}
\begin{figure}[t!]
\centering%
\includegraphics[width=0.8\textwidth]{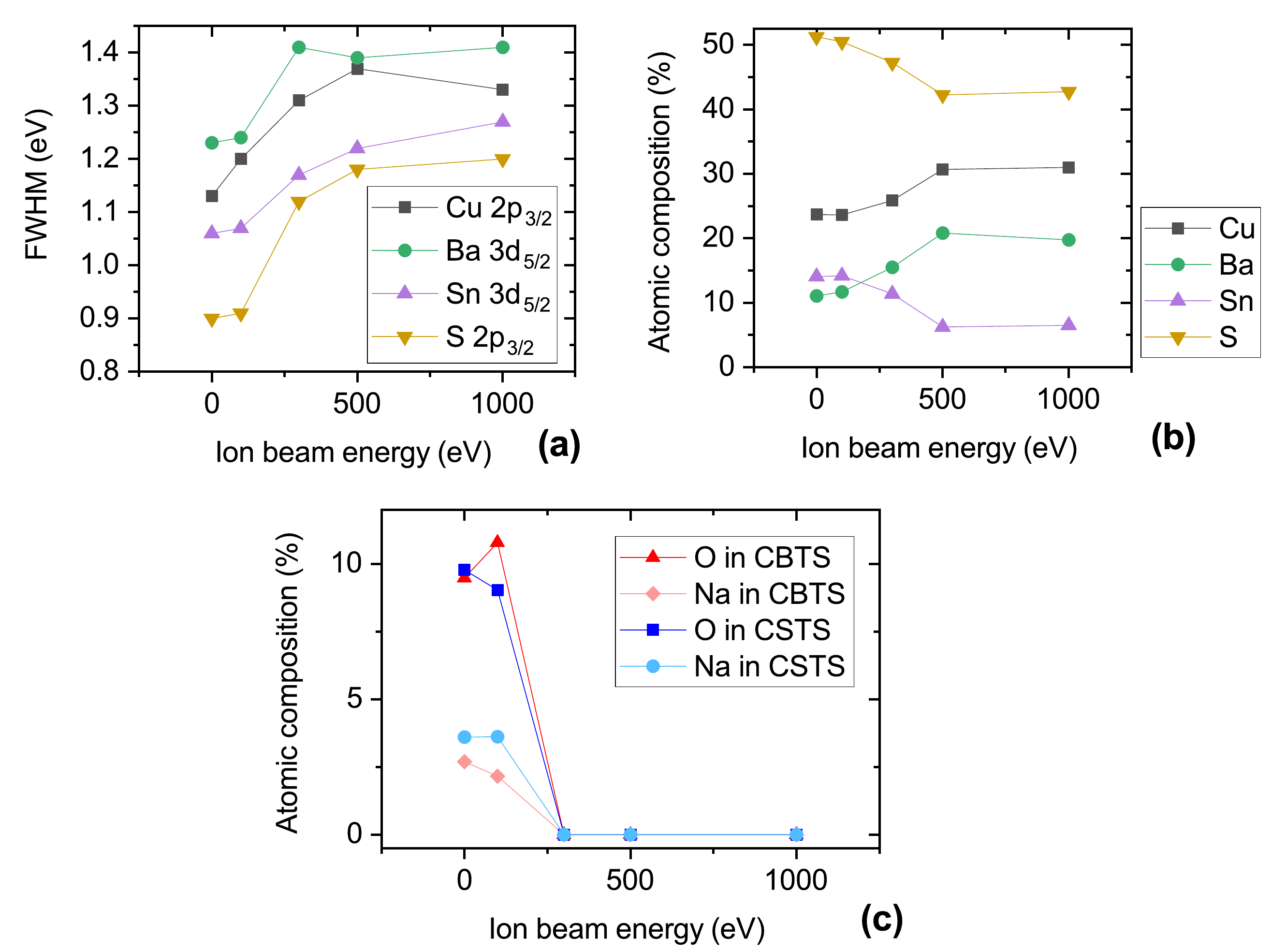}
\caption{(a): Full width at half maximum of various XPS core level peaks as a function of the energy of the Ar$^+$ ion beam used to clean the surface of the sample. (b): XPS-measured atomic composition of the CBTS surface (O and Na excluded) as a function of ion beam energy. Note the preferential loss of SnS already at low energies. (c): XPS-measured O and Na atomic composition in CBTS (expressed as fraction of the total Cu + Ba +Sn + S + O + Na composition) as a function of ion beam energy. For ion beam energies above 300~eV, both elements are below the detection limit.}
\label{fig:composition}
\end{figure}

\renewcommand{\thefigure}{S4}
\begin{figure}[t!]
\centering%
\includegraphics[width=0.8\textwidth]{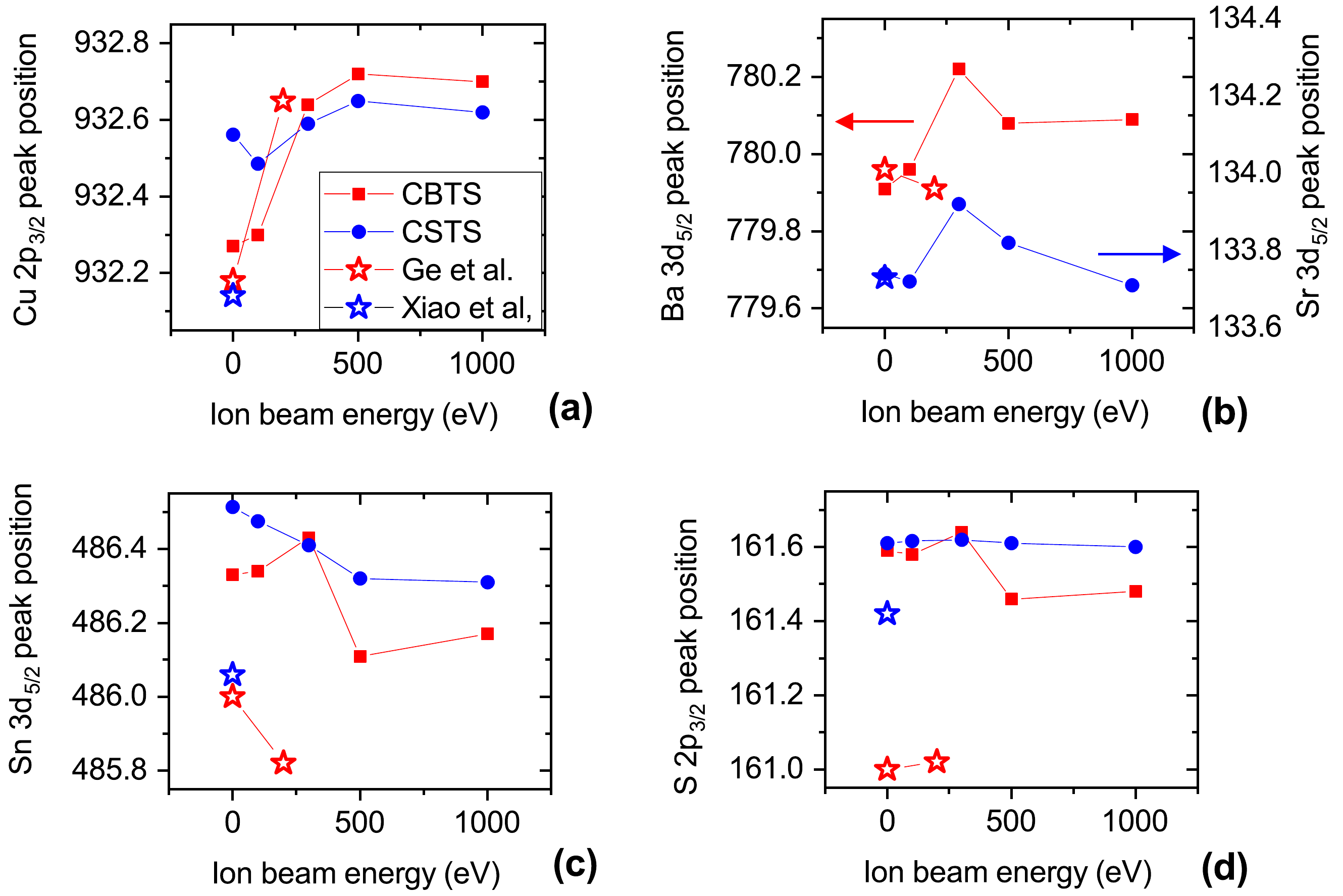}
\caption{Energy position of the main core level peaks of CBTS and CSTS as a function of the energy of the Ar$^+$ ion beam used to clean the surface of the sample. Red (blue) symbols indicate data points for CBTS (CSTS) films. The data points from the present study are shown as full squares and circles. Data points from previous studies (Ref.~\onlinecite{Ge} for CBTS, Ref.~\onlinecite{Xiao} for CSTS) are shown as empty stars.}
\label{fig:core_levels}
\end{figure}

\clearpage
\section*{Supplementary references}
\providecommand{\newblock}{}

\end{document}